\documentclass[12pt,preprint]{aastex}
\def\hho  {H$_2$O}
\def\HHO  {H$_2$O}
\def\kms  {\ifmmode {{\rm km~s}^{-1}} \else {km~s$^{-1}$} \fi}
\def\kmsperyr  {\ifmmode {{\rm km~s}^{-1} {\rm yr}^{-1}} 
                \else {km~s$^{-1}$ yr$^{-1}$} \fi}
\def\kmspmpc  {\ifmmode {{\rm km~s}^{-1} {\rm Mpc}^{-1}} 
                \else {km~s$^{-1}$ Mpc$^{-1}$} \fi}

\def\uas  {$\mu$as}
\def\etal {et al.~}
\def\eg   {e.g.~}
\def\ie   {i.e.~}
\def\UGC  {UGC~3789}
\def\NGC  {NGC~4258}
\def\Vlsr {\ifmmode {V_{\rm LSR}} \else {$V_{\rm LSR}$} \fi}
\def\Vhelio {\ifmmode {V_{\rm Helio}} \else {$V_{\rm Helio}$} \fi}
\def\Ho   {\ifmmode {H_0} \else {$H_0$} \fi}
\def\Msun {\ifmmode {M_\odot} \else {$M_\odot$} \fi}
\def\uvdata{({\it u,v})-data}

\def\didr {\ifmmode {\partial i/\partial r} \else {$\partial i/\partial r$} \fi}
\def\dpdr {\ifmmode {\partial p/\partial r} \else {$\partial p/\partial r$} \fi}
\def\didrtwo {\ifmmode {\partial^2 i/\partial r^2} \else {$\partial^2 i/\partial r^2$} \fi}
\def\dpdrtwo {\ifmmode {\partial^2 p/\partial r^2} \else {$\partial^2 p/\partial r^2$} \fi}
\def\domegadr{\ifmmode {\partial\omega/\partial r} \else {$\partial\omega/\partial r$} \fi}

\def\chisq{\ifmmode {\chi^2} \else {$\chi^2$} \fi}

\shorttitle{\UGC} 
\shortauthors{Reid \etal}

\begin{document}

\title{The Megamaser Cosmology Project: IV. \\
       A Direct Measurement of the Hubble Constant from \UGC}

\author{M. J. Reid\altaffilmark{1}, J. A. Braatz\altaffilmark{2},
        J. J. Condon\altaffilmark{2}, K. Y. Lo\altaffilmark{2}, 
        C. Y. Kuo\altaffilmark{2}, C. M. V. Impellizzeri\altaffilmark{2} \&
        C. Henkel\altaffilmark{3,4}}

\altaffiltext{1}{Harvard-Smithsonian Center for
   Astrophysics, 60 Garden Street, Cambridge, MA 02138, USA}
\altaffiltext{2}{National Radio Astronomy Observatory,
   520 Edgemont Road, Charlottesville, VA 22903}
\altaffiltext{3}{Max-Planck-Institut f\"ur Radioastronomie, 
   Auf dem H\"ugel 69, 53121 Bonn, Germany}
\altaffiltext{4}{King Abdulaziz University, P.O. Box 80203, Jeddah, Saudi Arabia}

\begin{abstract}
In Papers I and II from the Megamaser Cosmology Project, we reported 
initial observations of \HHO\ masers in an accretion disk of a
supermassive black hole at the center of the galaxy \UGC, which gave an 
angular-diameter distance to the galaxy and an estimate of \Ho with 
16\% uncertainty.  We have since
conducted more VLBI observations of the spatial-velocity structure of
these \HHO\ masers, as well as continued monitoring
of its spectrum to better measure maser accelerations.  These more extensive
observations, combined with improved modeling of the masers in the accretion
disk of the central supermassive black hole, confirm our previous 
results, but with significantly improved accuracy. 
We find $\Ho = 68.9\pm7.1$ \kmspmpc; this estimate of \Ho is independent of 
other methods and is accurate to $\pm10$\%, including sources of systematic error.
This places \UGC\ at an angular-diameter distance of $49.6\pm5.1$~Mpc, with a central 
supermassive black hole of $(1.16\pm0.12)\times10^7$~\Msun.   
\end{abstract}


\section{Introduction}
Measurements of the expansion history of the universe, $H(z)$, play a 
fundamental role in our understanding of cosmological evolution and its 
far-reaching physical implications, including addressing the nature of dark 
energy, the curvature of space, the masses of neutrinos, and the number of 
families of relativistic particles.  While detailed measurements of the 
cosmic microwave background (CMB, \eg  \citet{Komatsu:11}) allow us to measure 
linear scales at redshift $z\approx1100$, this mostly provides information at a 
single epoch and when the influence of dark energy was still negligible.
Complementary information from later cosmic times are therefore essential. 
This information can come from type Ia supernovae, galaxy clustering, gravitational lensing, 
and baryon acoustic oscillations (\eg  \citet{Riess:11,Bonamente:06,Suyu:10,BReid:10}).

All of these data provide distances and linear scales at significant redshifts. 
However, it is the local universe where dark energy is dominant. Thus the
local Hubble constant, $\Ho$, provides the largest ``lever arm'' with respect 
to the CMB for constraining the time dependence of the equation of state of 
dark energy. The use of
Cepheids (\eg \citet{Freedman:01,Sandage:06}) as well as the combined use of 
Cepheids and Type Ia supernovae (\eg \citet{Riess:11}) has traditionally 
dominated determinations of $\Ho$.  What is missing, however, are direct geometric
distance estimates that do not require a complex and uncertain ladder of
calibration of ``standard candles.'' 

Direct geometric distance measurements to water masers in nuclear regions 
of galaxies that are well into the Hubble flow (roughly $>30$~Mpc distant) 
provide a promising new and independent method for refining the value of \Ho.  
Observations of water masers in accretion disks within $\sim0.1$~pc of a 
galaxy's central supermassive black hole have been used to measure 
angular-diameter distances to galaxies, independently of other techniques that 
often rely on standard candles.  Very Long Baseline Array (VLBA) 
observations of the \hho\ masers in the nearby Seyfert 2 galaxy \NGC\ 
established the technique and provided an accurate, angular-diameter distance 
of $7.2\pm0.5$~Mpc to the galaxy \citep{Herrnstein:99}.  This galaxy is too 
close to permit a direct measurement of \Ho (\ie by dividing its 
recessional speed by its distance), since its uncertain peculiar velocity 
could be a large fraction of its recessional speed.  However, \NGC\ has proven 
extremely valuable as a solid anchor for the extragalactic distance scale 
\citep{Freedman:01,Riess:11}.  

Recently, the Megamaser Cosmology Project (MCP) reported a distance to 
\UGC, another galaxy with water masers in a nuclear accretion disk, of 
$49.9\pm7.0$~Mpc \citep{Reid:09,Braatz:10} (hereafter Papers I and II).  
This galaxy has a recessional velocity of $\approx3481$~\kms\ (relativistically 
corrected and referenced to the cosmic microwave background), 
which includes its peculiar velocity of $151\pm163$~\kms\  
based on galaxy flow models of \citet{Masters:06} and \citet{Springob:07}).  
Combining the recessional velocity and distance yielded $\Ho = 69 \pm 11$~\kmspmpc.   

Since \UGC\ is well into the Hubble flow it can provide a direct estimate of 
\Ho with a potential uncertainty as small as $\pm5$\%, limited by the 
uncertainty in its peculiar motion.   Therefore, since reporting our initial 
results, we have conducted additional observations of \UGC, in order to reduce 
the uncertainty of the \Ho estimate.  In total, we have now analyzed nine 
Very Long Baseline Interferometric (VLBI) 
observations, using the NRAO \footnote{The National Radio Astronomy 
Observatory is operated by Associated Universities, Inc., under a cooperative 
agreement with the National Science Foundation.} 
10-antenna VLBA, the 100-m Green Bank Telescope (GBT) and the 100-m Effelsberg 
\footnote{The Effelsberg 100-m telescope is a facility of the  
Max-Planck-Institut f\"ur Radioastronomie}
telescope.  The VLBI observations are reported in \S\ref{section:VLBI}.  
We have also extended our monitoring of changes in the water maser spectrum 
with monthly observations (except during the humid summer) with the GBT
for a period spanning 5.5 years.  
These spectra are documented in \S\ref{section:accelerations} and 
used to determine the accelerations of individual maser features.   
In \S\ref{section:fitting}, we describe a Bayesian approach for fitting a 
model of the accretion disk to these data.  The model allows for a warped 
disk with eccentric gas orbits and includes parameters for the central 
(black hole) mass and position, as well as for \Ho.

\section{VLBI Imaging} \label{section:VLBI}

VLBI observations were conducted at nine epochs (NRAO 
program codes BB227A, BB227B, BB242A, BB242I, BB242K, BB242L, BB242Q, BB261G, 
and BB261S) between December 2007 and April 2010.  
Six of the nine observations yielded maps with signal-to-noise
ratios degraded by factors of two or greater owing to the loss of one of the 
100-m telescopes (usually to poor weather) or the weakening of
the peak maser emission to levels where self-calibration (phase-referencing) 
using the maser emission was poor.  
The three epochs with excellent weather, antenna performance, 
and strong maser emission included BB227A 
(2006 December 10; reported in Paper I), BB242L (2008 December 12), and 
BB261G (2009 April 11).  Only results from these three epochs are reported here.

We observed with 16-MHz bands covering five frequencies, three in dual polarization
and two in single polarization.  In our early observations, these bands 
were centered at (optical definition) local standard of rest (LSR) velocities 
(and polarizations) of 3880.0 (LCP \& RCP), 3710.0 (LCP), 3265.0 (LCP \& RCP), 
2670.0 (LCP \& RCP), and 2500.0 (LCP)~\kms for BB227A; for the BB242L and 
BB261G, we shifted the center velocities of the fourth and fifth bands to 
2717.6 and 2513.6~\kms, in order to map new maser features not covered in
the original setup.  The data for each polarization of each band were
cross-correlated with 128 spectral channels, yielding channels separated
by 1.7 \kms.

Generally, the data were analyzed as described in Paper I.  
The final calibration step involved selecting a maser feature
as the interferometer phase-reference, and details of this procedure
varied among epochs depending on maser strength and interferometer coherence 
times.  The strongest maser feature in the spectrum usually
peaked at $\approx0.07$~Jy and was fairly broad.  
For BB227A, we averaged five spectral channels spanning an LSR velocity 
range of 2685 to 2692~\kms\ (\ie\ channels 52 to 56 from the blue-shifted
high-velocity band centered at $\Vlsr=2670$~\kms), adding together the data 
from both polarizations, and fitting fringes over a 1~min period.  
For BB242L, we averaged 13 spectral channels spanning an LSR velocity range of 
2682 to 2702~\kms\ (\ie\ channels 74 to 86 from the blue-shifted high velocity 
band centered at $\Vlsr=2717.6$~\kms) and fitted fringes over a 2~min period.  
For BB261G, we averaged seven spectral channels spanning an LSR velocity 
range of 2685 to 2696~\kms\ (\ie\ channels 78 to 84 from the blue-shifted
high velocity band centered at $\Vlsr=2717.6$~\kms), again with 2~min averaging.

After calibration, we Fourier transformed the gridded \uvdata\ 
to make images of the maser emission in all spectral 
channels for each of the five IF bands.  The images were deconvolved with the 
point-source response using the CLEAN algorithm and restored with a circular 
Gaussian beam with a 0.30~mas full-width at half-maximum (approximately matching
the geometric mean of the dirty beam).  The rms noise levels in channel maps
were $\approx1$~mJy.
All images appeared to contain single, point-like maser spots. 
We then fitted each spectral-channel image with an elliptical Gaussian 
brightness distribution in order to obtain positions and flux densities.

\begin{figure}[h]
\epsscale{0.9} 
\plotone{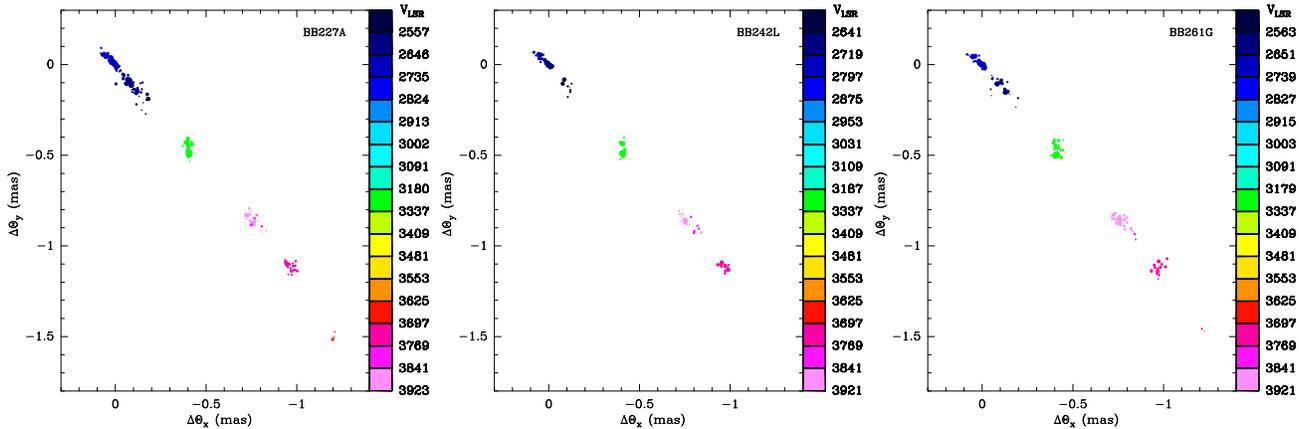} 
\caption{\footnotesize 
  Maps of the 22~GHz \hho\ masers toward \UGC\ constructed from VLBI data 
  using the VLBA, the GBT and the Effelsberg antennas for programs
  BB227A (2006 December 10), BB242L (2008 December 12) and
  BB261G (2009 April 11).  The LSR velocity of each maser spot is 
  indicated by the color bar on the right side of each panel. 
  \label{figure:maser_maps}
        }
\end{figure}

Spectral images for the three epochs are shown in Fig.~\ref{figure:maser_maps}
for maser spots stronger than 5 mJy.  (Note, that in Paper I, we displayed
only maser spots stronger than 10 mJy.)
These images show nearly identical spatial-velocity patterns of
maser emission.  This is expected as material in near-circular orbit in
the accretion disk at the average masing radius has an orbital period of $\sim1000$~yr
and thus rotates by less than $1^\circ$ of disk azimuth between the first
and last observation.   Most of the spatial scatter in the images is from 
measurement uncertainty, primarily from signal-to-noise limitations,
of $\approx25$~\uas\ for weaker features of $\approx10$~mJy.  

\section     {GBT Spectral Monitoring and Acceleration Fitting} \label{section:accelerations}

Paper II describes the procedures used to monitor the \HHO\ maser spectrum
of \UGC\ from January 2006 to March 2009 with the GBT.   We continued these 
observations through June 2011.  Typically, observations were conducted at
monthly intervals for nine months per year, avoiding the summer when 
atmospheric water vapor precluded sensitive observations.  These spectra were 
divided into 6 yearly blocks and analyzed to determine velocity drifts 
(accelerations) of individual maser features.   

Measurement of the acceleration of systemic-velocity masers is key to
determining the angular-diameter distance of the galaxy.  In principle,
accelerations are straight-forwardly obtained from the linear drift
in peak velocity with time.  In practice, however, blending of adjacent
spectral features, coupled with modest signal-to-noise ratio observations,
makes this difficult.   We applied two methods described below to measure accelerations: 
method-1 to obtain the best acceleration estimates and method-2 to check for possible 
fitting biases owing to initial parameter values.

\subsection  {Method--1} \label{subsection:method_1}

In Paper II we measured accelerations in two steps.  First, we identified 
spectral peaks in individual spectra ``by eye'' and tracked 
these peak velocities over time to obtain preliminary accelerations.
Then, we modeled the spectral flux density as a function of velocity and
time as the sum of a number of Gaussian spectral lines, whose center
velocity changes linearly with time.  We repeated this approach with
the extended spectral monitoring data.

Parameters for individual Gaussian components included the amplitudes 
(one per observational epoch), the line width, and the center velocity 
(at a reference epoch) and its (linear) change in velocity over time.  Initial
values for maser component velocities and accelerations were set based on 
the ``by eye'' values; line widths were initially set at 2.0~\kms\ and
assumed not to vary over the $\approx9$ months of observations being analyzed. 
Initial values for maser component amplitudes for each observation were set 
automatically at the flux density of each spectrum at the velocity determined
by the initial central velocity and acceleration parameters.

The parameters were adjusted by a sequence of least-square fitting  
(to minimize the sum of the squares of the weighted post-fit
residuals or $\chi^2$).
In the first fitting step, only the amplitudes were adjusted.  Next,
the amplitudes, central velocities and line widths were adjusted,
using parameter values derived from the first fits.
Finally, all parameters (including accelerations) were allowed to vary.
While this least-squares approach works well and gives 
accelerations that largely agree with those visually evident in the spectra, 
it is a time-consuming process and dependent on somewhat subjectively 
determined initial parameters. 

\subsection  {Method--2} \label{subsection:method_2}

In order to avoid setting the initial velocities and acceleration parameters  
by eye, we developed an alternative analysis method.
This method involved randomly choosing initial values followed by least-squares 
fitting and evaluation of the quality of each fit.
We repeated this process, with independently chosen initial parameter values, 
about 100 times and recorded the four fits with the lowest values of $\chi^2$ 
per degree of freedom.  In detail, for each trial solution, we assigned the 
center velocity, $V_n$, of the n$^{th}$ spectral component by selecting a 
velocity offset randomly, from a Gaussian distribution with a mean of 
2.0~\kms\ and a standard deviation of 0.7~\kms, and adding this offset to the 
central velocity of the (n-1)$^{th}$ component.  We started at the low end of 
the velocity ``window'' being fitted and continued until we reached the high 
end; this allowed different trial fits to have different numbers of velocity 
components.  

Since the change in component acceleration with velocity is generally small, 
rather than set accelerations independently for each component as in
method--1, we set a single acceleration and its velocity derivative 
(2 parameters) for the entire velocity window under consideration.   
This minimizes the number of free parameters, but requires setting small 
velocity windows over which component accelerations are nearly constant.  
Specifically, we assigned an initial acceleration, $A_n$, to the n$^{th}$ 
velocity component given by $A_n = A_c + (dA/dV)(V_n - V_c)$, where $V_c$ is 
the center of the velocity window, $A_c$ is the average acceleration over the 
fitting window and $dA/dV$ allows for a linear change in acceleration with 
velocity.  Values of $A_c$ were chosen randomly from a Gaussian distribution, 
whose mean was estimated from the fits described in method--1 and 
whose standard deviation, $\sigma_A$, was one-third of that mean.  
Values of $dA/dV$ were chosen in a similar random manner from a distribution 
with zero mean and standard deviation of $\sigma_A/1~{\rm yr}$.
The combined effect of the large standard deviation for accelerations 
and allowance for a linear change in acceleration with velocity resulted
in broad sampling of initial acceleration parameter space.

\subsection  {Acceleration Fitting Results} \label{subsection:acc_fitting}

For systemic velocity features, we used three velocity windows for acceleration 
fitting: 3230--3280, 3286--3316, and 3350-3370 \kms, since emission outside 
of these windows was generally absent or very weak ($<5$~mJy).
Systemic velocity maser features in \UGC\ typically have lifetimes (with flux 
densities $>5$~mJy) of $\sim12$ months.  Therefore, we fitted accelerations 
separately to groups of $\approx9$ consecutive monthly spectra (from fall 
through spring) covering 5.5 years and centered at $\approx2008.8$.
These observations well-covered the three high-quality VLBI imaging observations
discussed in this paper, whose mean epoch was $\approx2008.4$.   

\begin{figure}[h]
\plotone{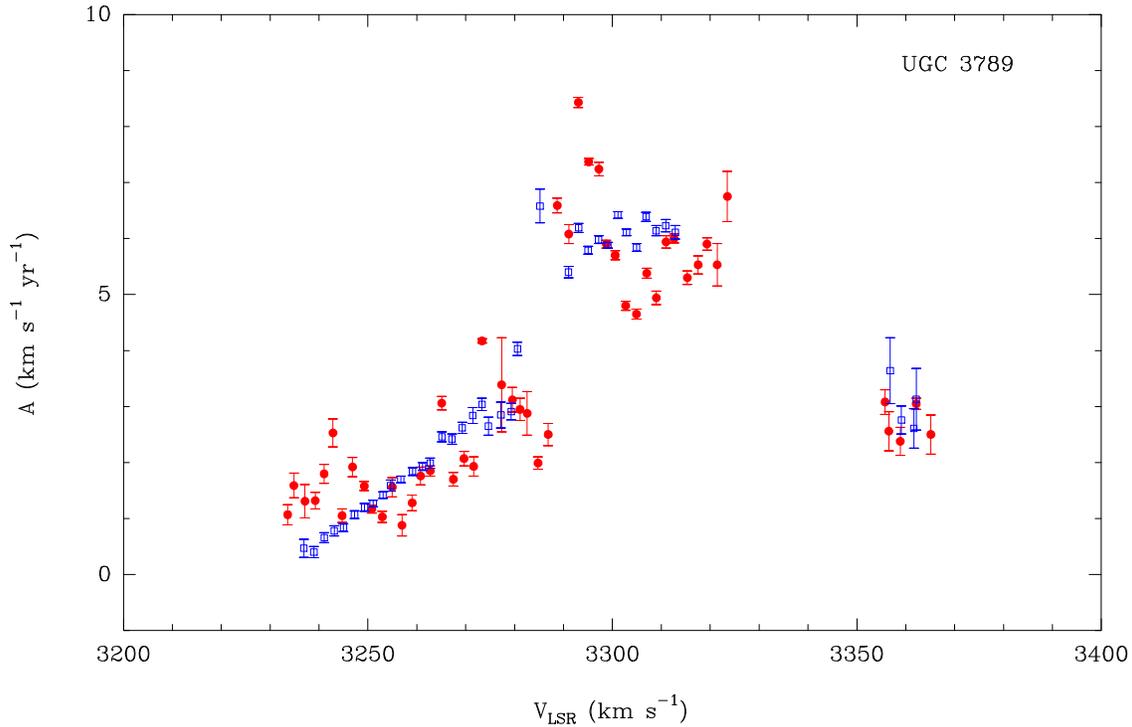} 
\caption{\footnotesize 
  Acceleration measurements as a function of LSR velocity for
  systemic velocity maser features.  {\it Filled red circles}
  are results from method--1 (see \S\ref{subsection:method_1}) 
  and {\it open blue squares} are from method--2
  (see \S\ref{subsection:method_2}).  
  \label{figure:systemic_accelerations}
        }
\end{figure}

Yearly acceleration measurements generally were consistent with a single
value, although for some velocity ranges there was
considerable scatter (up to about $\pm30$\% for masers near 3290~\kms).
Because of this, and the expectation of small changes in acceleration
over our observing period (as masing clouds move by less than $1^\circ$ 
of disk azimuth), we velocity binned and averaged the fitted accelerations 
from the 6 ``yearly'' groups of spectra.  Variance weighted averages of 
acceleration as a function of velocity for the systemic features are plotted 
Fig.~\ref{figure:systemic_accelerations} for both fitting methods outlined above.  

The results for method-1 are plotted with filled (red) circles in
Fig.~\ref{figure:systemic_accelerations}.
There is moderate scatter among the accelerations within each 
velocity window, and this scatter greatly exceeds that expected from the
formal acceleration uncertainties.  
The greatest variations occur in the velocity range 3279 -- 3330~\kms;
we can bound this variation by assuming a constant acceleration over this
velocity range and calculating a standard deviation of 1.6~\kmsperyr.
We suspect that some of the variation is real and intrinsic to the source, but
also that some of the variations may originate in the fitting process,
owing to multiple velocity-blended components and using data with only moderate 
signal-to-noise ratios. 
%However, the average acceleration values in each window are well determined and, 
%especially, in the 3230--3280 window, there is a smooth change in 
%acceleration with velocity.  

In order to provide an independent check on the results from method-1, we 
re-fitted the spectra using method-2.    
For each velocity window and each ``yearly'' time group, we 
selected the four trial fits with the lowest $\chi^2$ values.  We then binned all
results in 2~\kms\ bins and calculated a weighted average acceleration for
each velocity bin.  These results are plotted in 
Fig.~\ref{figure:systemic_accelerations} with open (blue) squares.
There is very good agreement between the two methods; the differences in average 
accelerations in each velocity window are small; unweighted means for methods-1 and -2 are
1.7 and 1.7 ($\pm0.2$) \kmsperyr\ for the velocity range 3230--3278 \kms, 
5.3 and 5.9 ($\pm0.4$) \kmsperyr\ for the velocity range 3279--3330 \kms, and 
2.7 and 3.0 ($\pm0.3$) \kmsperyr for the velocity range 3350--3370 \kms.   
The small differences between
the slopes of $A$ vs. $V_{LSR}$ for the two methods are within statistical 
uncertainties after accounting for the high correlation among accelerations within 
a window for method-2.
We conclude that the accelerations given by the solid (red) points in 
Fig.~\ref{figure:systemic_accelerations} are near optimum values and are
not significantly biased by the choice of initial parameter values in the 
first step of method-1.

Acceleration measurements for high velocity features are considerably less 
complicated as these features are not highly blended and have small changes 
in velocity over time.  Therefore, we only used method--1 (described in 
\S\ref{subsection:method_1}) for these features, and fitted accelerations are
given in Table~\ref{table:data}.

\section {Modeling the Accretion Disk \& Estimating \Ho} \label{section:fitting}

The position-velocity measurements from the three VLBI maps were combined by
binning the velocities in 2.0~\kms\ wide bins (comparable to maser linewidths) 
and calculating variance weighted positions and velocities.
We associated these positions and velocities with maser feature accelerations,
from the GBT time monitoring of spectra, by choosing the VLBI velocity
closest to that of an acceleration fit.  As there are fewer acceleration
fits than VLBI position-velocity measurements, not all VLBI measurements
have corresponding accelerations.  For features lacking acceleration measurements 
we use only the position-velocity data when modeling.  The velocity, position and 
acceleration values (when available) are given in Table~\ref{table:data} and used 
to model the accretion disk.  

Rather than use formal fitting uncertainties, which tend to be optimistic for high 
signal-to-noise data, we adopted more realistic ``error floors'' for the data 
uncertainties of $\pm0.01$ mas for positions, $\pm1.0$ and $\pm0.3$ \kms\ for the 
velocities of systemic and high-velocity maser features, respectively, and 
$\pm0.57$ \kmsperyr\ for accelerations.   
Error floors were added in quadrature to formal fitting 
uncertainties.  For example, systematic errors of $\approx0.01$~mas
between systemic and high velocity features can be caused by an error in
the absolute position of the reference maser spot of a few mas \citep{Argon:07}.
Using error floors allows not only for systematic uncertainty not captured by
formal estimates, but also for slight incompleteness in the model of the 
accretion disk, such as could come from unmodeled spiral structure.
For example, there are indications of slight departures from perfect Keplerian gas 
orbits about a dominant central mass in the nearby, well-studied, disk 
of NGC~4258 at levels of $\approx0.5$~\kmsperyr\ for systemic feature
accelerations \citep{Humphreys:08}.  
In \S\ref{subsection:weighting}, we explore the sensitivity
of the fitted parameters to changes in the magnitudes of these error floors.

Conceptually, were the masers in a perfectly
thin and flat accretion disk orbiting circularly about a point mass and 
viewed edge-on, the position-velocity data for the high velocity features 
would trace a Keplerian profile (\ie $V=\sqrt{GM/r}$), where $G$ is the 
gravitational constant, $M$ is the mass of the central black hole, and $r$ is 
the distance of a masing cloud from the black hole.  The symmetry of the 
approaching and receding features can be used to precisely locate the central 
black hole both in position ($x_0,y_0$) and velocity ($V_0$).  
Thus, for high velocity features that are in the plane of the sky, the VLBI 
image directly gives a feature's angular radius, $\theta = r / D$, where $D$ is the 
distance to  the galaxy.  Systemic velocity masers in front of the black hole, moving 
transversely on the sky, display a change in velocity over time ($A$) as 
they orbit the central mass.  Features in a thin annulus at radius, $r$, 
will be observed to accelerate at $A = V^2/r$.  Thus, from position-velocity
and acceleration measurements, one can estimate  $D = V^2/A\theta$ and then 
$\Ho \approx V_r / D$, where $V_r$ is the recessional velocity.  
Note that proper motions of systemic-velocity masers
can also be used to estimate distance, and \citet{Herrnstein:99} showed that
the proper motion distance agreed with the acceleration distance estimate
for the nearby galaxy NGC~4258.  However, proper motions decrease in magnitude
with source distance and are generally far less accurate than radial 
acceleration measurements based on changing Doppler shifts (which are distance 
independent).

In practice, accretion disks are somewhat more complicated
than the idealized case just described.  The disks are not precisely
edge-on (although strong maser amplification prefers disk spin axes to be 
within about $5^\circ$ of the plane of the sky (\eg \cite{Watson:94}), 
disks are often somewhat warped, 
and gas may be on slightly eccentric orbits.  Also, galaxies typically have peculiar 
motions of hundreds of~\kms\ with respect to a pure Hubble flow (\eg \cite{Masters:06}).
These complications are best addressed by constructing a model for maser orbits in 
the disk and adjusting the model parameters to best match the observations.

We modeled the system with up to 13 global parameters.  The central black hole has a
mass, $M$, is located at ($x_0,y_0$) on the sky (relative to the reference
maser feature) and has a line-of-sight velocity, $V_0$, in the CMB rest frame.  
For \UGC, $V_0 = \Vlsr + 60$~\kms.   Following \citet{Herrnstein:99}, 
we model the disk warp by a change in inclination with radius, 
$i(r)=i_0 + (\partial i/\partial r)\delta r$, 
and position angle (defined east of north) with radius,
$p(r)=p_0 + (\partial p/\partial r)\delta r$,
where $\delta r = r - r_{ref}$ and $r_{ref}$ is a reference radius assigned to
the middle of the maser distribution.  
(Given the very small warping evident in the VLBI maps, we did not use second-order
warping terms.)
Our model can allow the masers to have eccentric orbits, with eccentricity $e$ and 
pericenter rotated in angle $\omega = \omega_0 + (\partial\omega/\partial r)\delta r$ 
with respect to our line of sight.
Finally, the galaxy (angular diameter) distance is calculated from \Ho, assuming its 
observed radial motion deviates from a pure Hubble flow by a peculiar 
velocity $V_p$ relative to the cosmic microwave background reference frame.  
The angular diameter distance is calculated rigorously from formulae of 
\citet{Hogg:99}, assuming cosmological parameters for matter density $\Omega_m=0.27$ 
and dark energy $\Omega_\Lambda=0.73$.

We chose to estimate $\Ho$ directly, rather than fit first for $D$, followed by
a second step to estimate $\Ho$.   There are several advantages in our approach.
Firstly, unlike NGC~4258, which is not sufficiently distant to be in the
Hubble flow and hence cannot be used to accurately estimate $\Ho$, for UGC~3789
our primary interest is $\Ho$ and not $D$.  Secondly, when solving for $D$, 
systemic feature accelerations, which are the least accurately measured
type of data, appear in the denominator ($D = V^2/A\theta$) and lead to an
asymmetric posteriori probability density function (PDF) and complicate the error 
estimation for $\Ho$.  Thirdly, we can directly incorporate uncertainty in
the galaxy peculiar velocity ($V_p$) into the uncertainty in \Ho\  through
the prior on $V_p$.

In addition to the global parameters, each maser feature requires two parameters,
its radius, $r$, and azimuth, $\phi$, to specify its location in the disk.
Initial values for $r$ and $\phi$ for high-velocity features were estimated
from the position data, assuming $\phi=90^\circ$ for red-shifted features and 
$\phi=-90^\circ$ for blue-shifted features.  Systemic velocity features were
assigned initial $r$ values based on the acceleration data and $\phi$ values 
based on velocity data.  All $(r,\phi)$ parameters were assigned flat priors 
and were adjusted in the fitting process.

We evaluated parameter {\it posteriori} PDFs with Markov 
chain Monte Carlo (McMC) trials that were accepted or rejected according to the
Metropolis-Hastings algorithm. All parameters except for one were given flat 
priors and hence could vary freely, constrained only by the difference between 
data and model.  The only parameter with a constraining prior was  
the galaxy's peculiar velocity, $V_p$; this 
comes from large scale gravitational perturbations \citep{Masters:06,Springob:07}
which are expected to cause observed velocities for UGC~3789 to be
lower by $151\pm163$ \kms\ with respect to the Hubble flow (\ie\  $cz = V_0 + V_p$).
All velocities quoted here are non-relativistic, optical definition, in the CMB frame;
the model and the data velocities (shifted from the observed LSR frame to the CMB 
frame by adding 60~\kms) were converted internally in the fitting program to 
relativistically correct values.
Initial runs indicated very little warping (as is evident in the VLBI images) and 
maser orbital eccentricities near zero.  Hence, for our ``basic model,'' 
we used only the first order warping parameters and assumed circular gas orbits.

We ran 10 ``burn-in'' stages, each with $10^6$ trials, in order to arrive at
near optimum parameter values and parameter step sizes.  Parameter step sizes
were iteratively adjusted after each burn-in stage so as to come from Gaussian 
distributions with similar widths as the anticipated {\it posteriori} PDFs 
(estimated from the burn-in McMC trials), multiplied by a global 
step-size factor.  This factor 
($\approx0.02$) was also adjusted in the burn-in stages to scale parameter steps 
so that an optimal Metropolis-Hastings acceptance rate near 23\% was achieved.
The widths of the anticipated {\it posteriori} PDFs
of the parameters and the global step-size factor, which together determine 
parameter step sizes, were taken from the last burn-in stage and then held constant 
for the final McMC trials.

After discarding the burn-in stage trials, we evaluated $10^7$
McMC trials to obtain final {\it posteriori} PDFs.  
The Pearson product-moment correlation coefficients from these trials are given
in Table~\ref{table:correlations}. 
The only correlation coefficient (r) with magnitude greater than 0.5 is between 
the Hubble constant and central black hole mass (r$_{\Ho,M}\approx-0.85$) 
(see discussion in \citet{Kuo:13}).  
In order to more optimally sample the PDFs for these 
parameters, we modified the McMC values for these parameters to be totally 
anti-correlated for half of the trials and uncorrelated for the remainder 
\citep{Gregory:11}.  

\begin{figure}[h]
\plotone{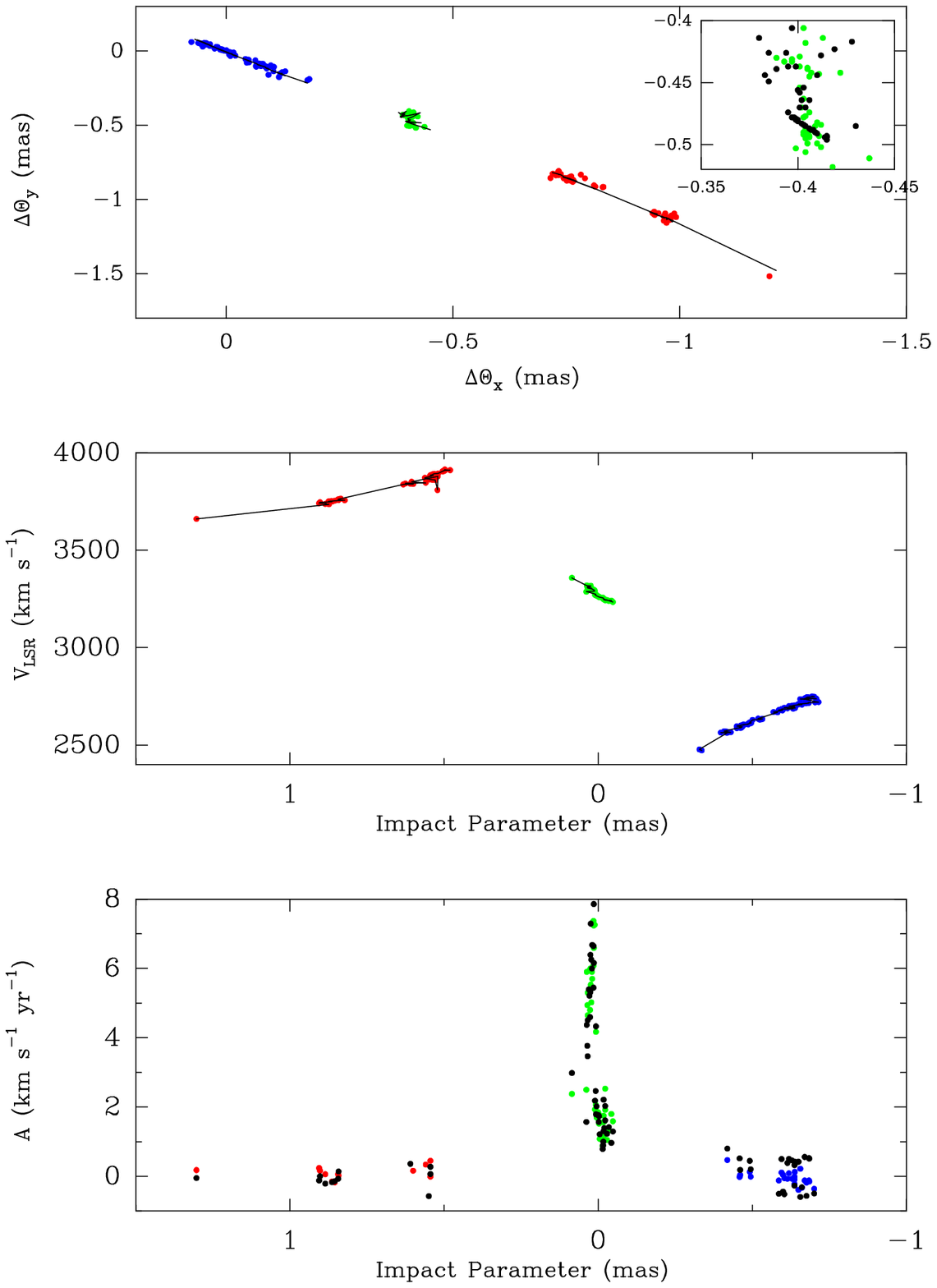} 
\caption{\footnotesize 
  Data (colored dots) and best-fit model (lines and black dots).
  {\it Top panel}: Positions on the sky.  Insert shows a blow-up of the 
  systemic-velocity masers.
  {\it Middle panel}: LSR velocity versus position along the disk.
  {\it Bottom panel}: Accelerations versus impact parameter.
  \label{figure:model_fit}
      }
\end{figure}

\begin{figure}[h]
\plotone{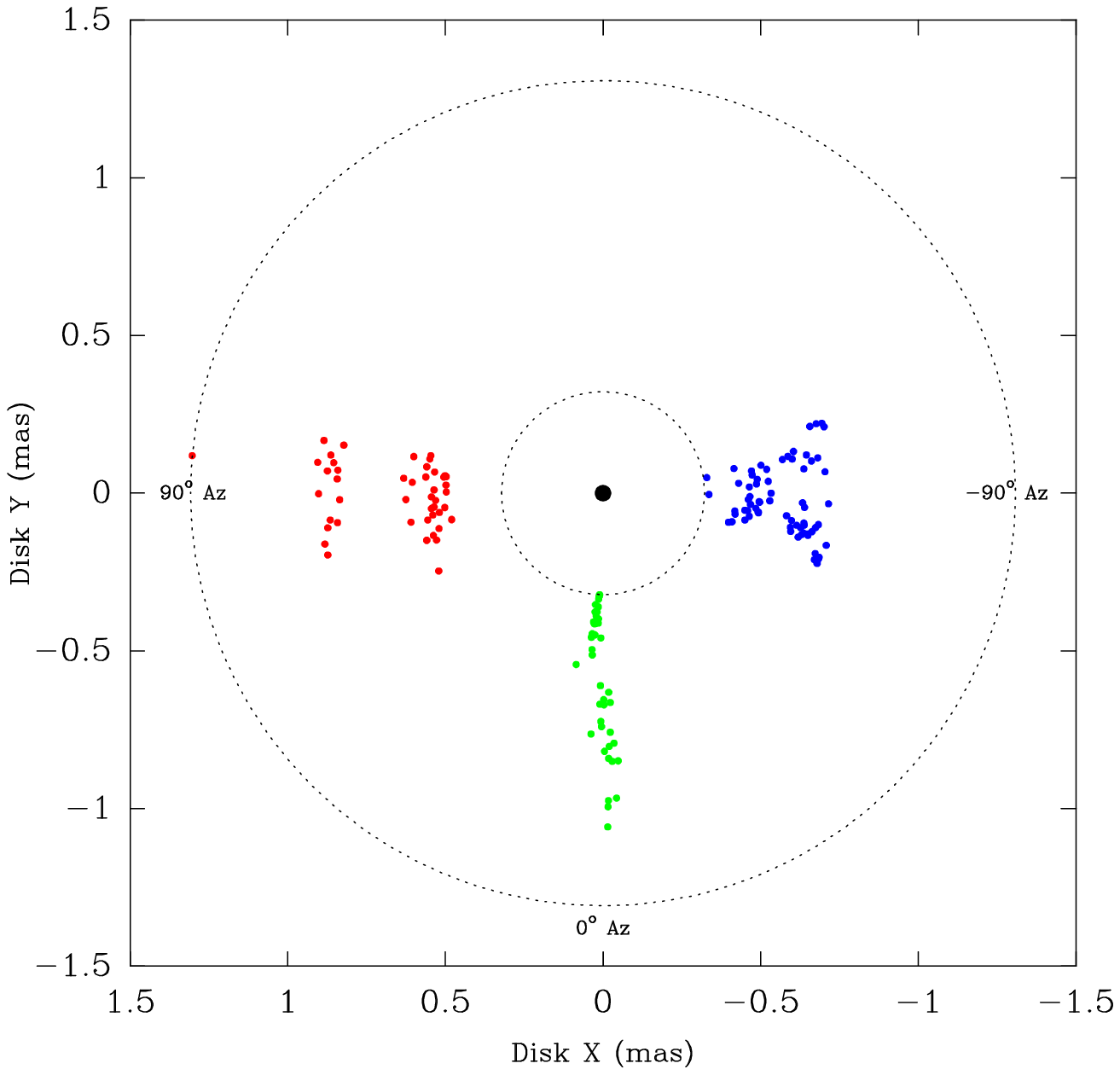} 
\caption{\footnotesize 
  Location of maser features projected in the plane of the
  accretion disk based on the best-fit model.  
  The location of the central black hole is shown with a {\it black filled circle} 
  at the origin.  The observer is at a large negative Y location.  Disk azimuth
  defined is as $0^\circ$ toward the observer. {\it Red and Blue} dots indicate the
  red-shifted (toward disk azimuth $\approx90^\circ$) and blue-shifted (toward disk azimuth 
  $\approx-90^\circ$) high-velocity masers, respectively.
  {\it Green} dots indicate the systemic velocity masers (toward disk azimuth 
  $\approx0^\circ$).  
  \label{figure:disk_plane}
      }
\end{figure}

From Bayes' theorem, the probability density of the parameters ($\rho$), 
given a model ($m$), data ($d$), and priors ($I$), is given by 
$$P(\rho|m,d,I) \propto P(d|,\rho,m,I) \times P(\rho|m,I)~~.$$
Since we assume Gaussianly distributed uncertainties for the data and priors, 
maximizing $P(d|\rho,m,I)\times P(\rho|m,I)$ is equivalent to minimizing 
$\chi^2_d + \chi^2_\rho$, where the subscripts $d$ and $\rho$ refer to the data and 
model parameters (given the priors), respectively.  The best-fitting trial gave 
$\chi^2_d = 1.50$ for 227 degrees of freedom.  In Fig.~\ref{figure:model_fit}, we 
show projections of the data with the best-fitting model superposed.   
The modeling also yields the $(r,\phi)$ coordinates of each maser feature
in the disk plane, which are displayed in Fig.~\ref{figure:disk_plane} (neglecting
the slight warping of the disk).  

Optimum values of the model parameters were
estimated from the {\it posteriori} PDFs marginalized over all  
other parameters.  They were generated from a total of $10^7$ McMC trials, 
obtained from 10 independent program runs starting with slightly different parameter 
values and new random number generator seeds.  Each run produced $10^7$ trials, 
but stored only every tenth trial (\ie\ ``thinned'' by a factor of 10). 
Parameter values given in Table~\ref{table:fitted_parameters} were produced
from binned histograms for each parameter and finding the sample median and 
$\pm34$\% ($\approx\pm1\sigma$) range.  Of some interest is the very
accurate estimate of the mass of the central black hole of 
$(1.16\pm0.12)\times10^7$~\Msun.

\begin{figure}[h]
\plotone{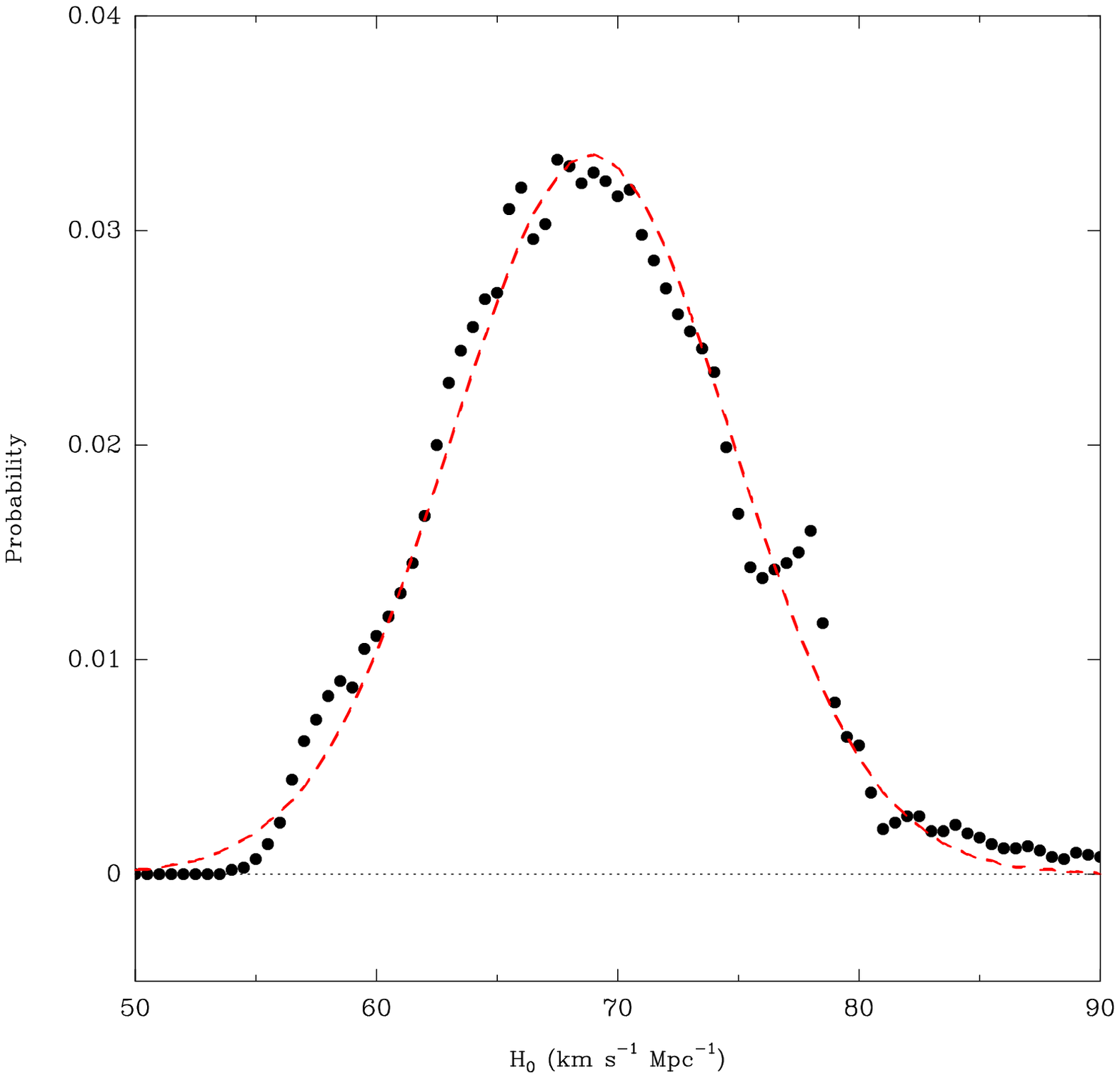} 
\caption{\footnotesize 
  Posteriori probability density function for the Hubble constant
  parameter (\Ho), marginalized over all other parameters.
  Superposed ({\it dashed red line}) is a Gaussian with $\sigma=5.8$ \kmspmpc.
  Scaling the Gaussian width by $\sqrt{\chi^2_\nu}$ yields
  our estimate of the Hubble constant of $\Ho = 68.9\pm7.1$~\kmspmpc.
 \label{figure:Ho_pdf}
      }
\end{figure}

\begin{figure}[h]
\plotone{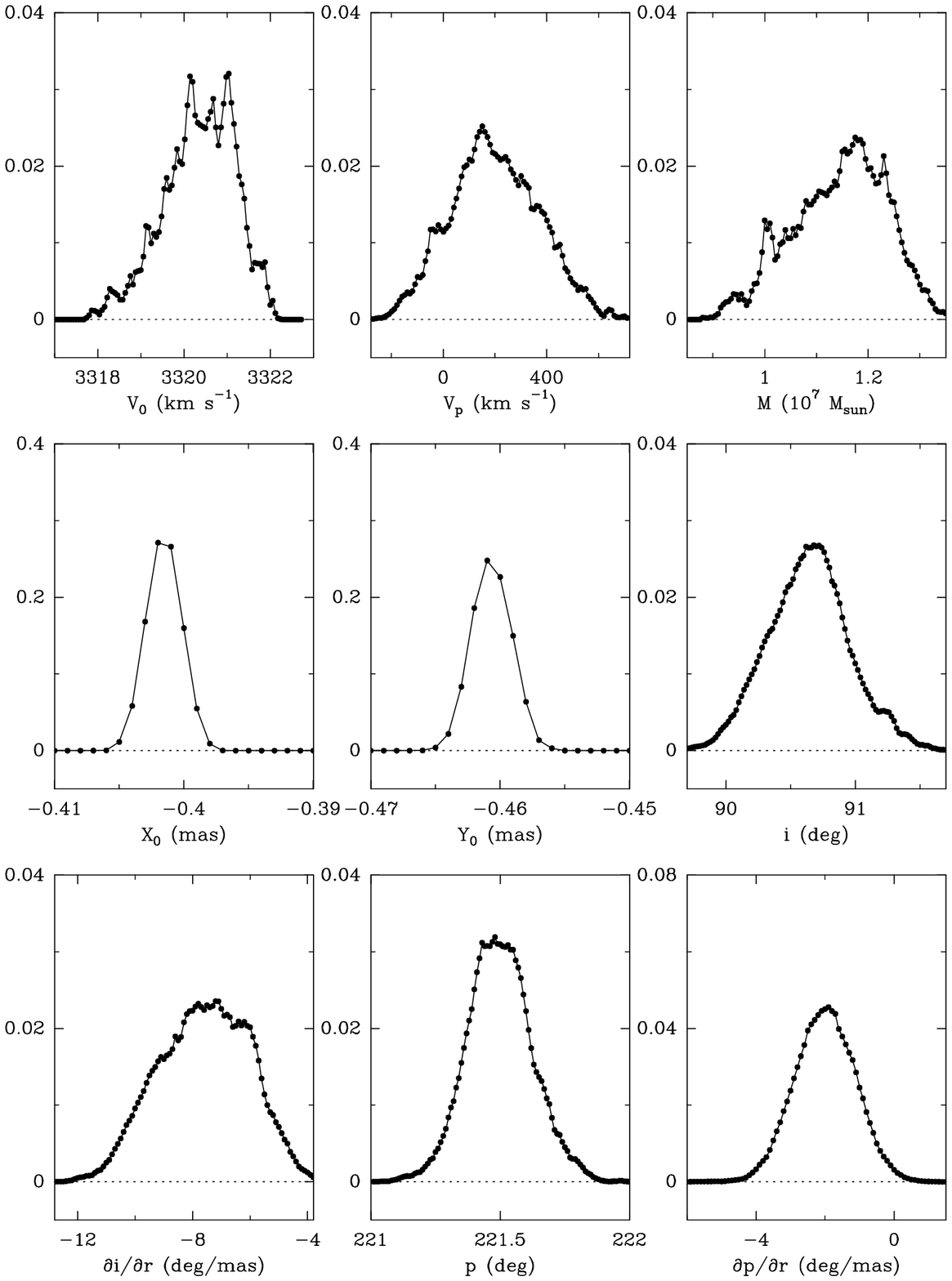} 
\caption{\footnotesize 
  Marginalized posteriori probability density functions for the nine global 
  parameters, excluding \Ho\ which is shown in Fig.~\ref{figure:Ho_pdf}.
 \label{figure:9_pdfs}
      }
\end{figure}

The binned {\it posteriori} PDF for the Hubble constant,
marginalized over all other parameters, is displayed in Fig.~\ref{figure:Ho_pdf},
and the marginalized distributions for the other 9 global parameters are
shown in Fig.~\ref{figure:9_pdfs}.  
The \Ho\ distribution can be well approximated by a Gaussian with a mean of $68.9$ \kmspmpc\ 
and a $\pm34$\% confidence range of $\sigma=\pm5.8$~\kmspmpc.
Since, the reduced $\chi^2_\nu = 1.50$, we conservatively inflate this uncertainty 
by a factor of $\sqrt{1.50}$ to $\pm7.1$~\kmspmpc.  This Hubble constant, 
coupled with the recessional velocity of $cz = V_o + V_p = 3466$~\kms 
(optical definition relative to the CMB, and corrected for a peculiar velocity of $151$~\kms), 
and assuming cosmological parameters for matter density $\Omega_m=0.27$ and dark
energy $\Omega_\Lambda=0.73$, places \UGC\ at an angular-diameter distance of $49.6\pm5.1$~Mpc.
This corresponds to a luminosity distance of $50.8\pm5.2$~Mpc.

\subsection {Sensitivity of $\Ho$ to Data Weighting} \label{subsection:weighting}

We tested the sensitivity of our estimate of $\Ho$ to significant ($>30$\%) 
changes in the error floors applied to the data set of Table~\ref{table:data}.
Changing the positional error floors from 0.010 to 0.005 or 0.015~mas changed
estimates of $\Ho$ by less than 1~\kmspmpc.  Varying
the acceleration error floor from 0.57 to 0.27 or 0.87~\kmsperyr\ produced
similarly small changes in $\Ho$.  Decreasing the velocity error floors
for the systemic and high-velocity maser features from 1.0 and 0.3 ~\kms,
respectively, to 0.7 and 0.2~\kms\ also yielded insignificant changes in $\Ho$.  

In our tests, the only sensitivity of $\Ho$ to changes in error floors
occurred when increasing the velocity error floors for systemic and
high-velocity maser features from 1.0 and 0.3~\kms, respectively, by 30\%\  
to 1.3 and 0.4~\kms.  This resulted in estimates 
of $\Ho$ reduced by 4~\kmspmpc.  This small sensitivity likely comes from
down-weighting the velocity information in the high-velocity masers.  This
information is critical to the method as it provides direct and strong constraints
for the location ($x_0,y_0$) and velocity ($V_0$) of the central black hole.
Down-weighting this data, by increasing its uncertainty, requires the program 
to use weaker, indirect information in the position/acceleration data to determine 
these parameters.

\subsection {Secondary $\chisq$ Minima} \label{subsection:chisq}

In preliminary attempts to fit the data, we used a prior constraint for 
$\Ho$ of $72\pm12$~\kmspmpc, probably at least double the current uncertainty
in the Hubble constant.   However, we ultimately dropped this constraint
in favor of a flat prior on $\Ho$, allowing us to arrive at an estimate
of $\Ho$ from UGC~3789 data alone that is independent of prior knowledge.
We then tested the sensitivity of the estimates of $\Ho$ to the initial values.  
Starting with $\Ho$ values as high as 80~\kmspmpc, we found fitted values 
returning close to our base result near 70~\kmspmpc.  However, starting $\Ho$ 
below 60~\kmspmpc, we found a stable fit with $\Ho = 59$~\kmspmpc.

The $\Ho=59$~\kmspmpc\ result likely comes from a secondary minimum in $\chisq$
space.  For the fit returning $\Ho=59$, $\chisq = 370.9$ for 227 degrees of freedom.
This can be compared with the better $\chisq = 360.0$ (for the same 227 
degrees of freedom) for our basic fit with $\Ho=68.9$~\kmspmpc.  Because
the $\Ho=59$~\kmspmpc\ fit produces a significantly larger $\chisq$, we
exclude this trial.

\subsection {Eccentric Gas Orbits} \label{subsection:eccentricity}

In order to assess the sensitivity of $\Ho$ estimates to the 
assumption of circular gas orbits used in our basic model, we re-fit the data with 
a more general model with 3 additional parameters ($e$,$\omega$ and $\domegadr$) 
that allow eccentric gas orbits with pericentric angle changing linearly with 
radius in the accretion disk.  Best-fit values for eccentricity were very small,
$0.025\pm0.008$, with pericenter at $-60^\circ\pm20^\circ$ disk azimuth.  Such a
small eccentricity has a negligible effect on the other parameter estimates.

\section {Dark Energy Constraints}

Direct measurements of $\Ho$, such as from \UGC, are especially important for constraining 
the equation of state of dark energy, $w$, since the effects of dark energy are 
greatest at the present epoch.  Our estimate of $\Ho$ can
be combined with results from the Wilkinson Microwave Anisotropy Probe (WMAP) mission to
tighten constraints on the dark energy equation of state, $w$, {\it independent} of other 
methods such as using SN~Ia or baryon acoustic oscillations \citep{Eisenstein:05}.  

\begin{figure}[h]
\plotone{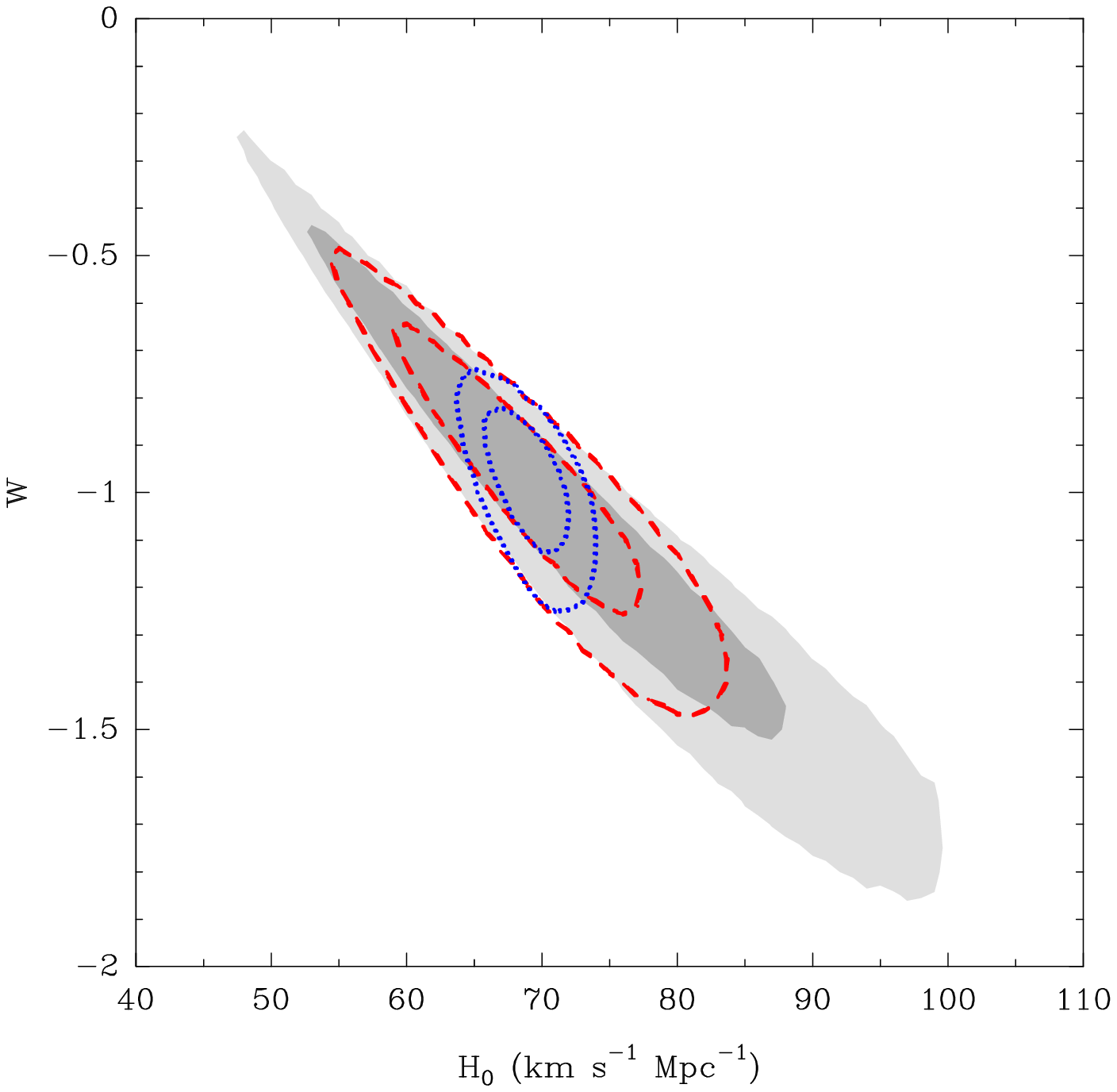} 
\caption{\footnotesize 
  2-D probability density functions for a (constant) equation of state
  of dark energy ($w$) and \Ho.  {\it Grey scale contours} come from 
  WMAP7.2 results.  {\it Red dashed contours} combine the WMAP probability
  density function and the constraint that
  $\Ho=68.9\pm7.1$~\kmspmpc\ from the results of \UGC\ presented in this
  paper.  {\it Blue dotted contours} anticipate an improved constraint of
  $\Ho=68.9\pm2.1$~\kmspmpc\ from the results of $\approx10$ galaxies
  like \UGC, the goal of the Megamaser Cosmology Project.
  Contours enclose 68\% and 95\% probabilities. 
 \label{figure:w_vs_Ho}
      }
\end{figure}

Fig.~\ref{figure:w_vs_Ho} shows 2-dimensional PDFs for $w$ 
and \Ho with 95\% and 68\% confidence contours.  The grey-scale contours were generated 
by binning the parameter values from Markov chains 
(wmap\_wcdm\_sz\_lens\_wmap7.2\_chains\_norm\_v4p1.tar.gz)
from the WMAP 7-year data (processed with WMAP version 4.1, RECFAST version 1.5
and modeled with a constant-$w$, $\Lambda$CDM model that incorporates the 
effects of the SZ effect and gravitational lensing).   
Fitting a Gaussian to the marginalized 1-dimensional
PDF for $w$ yields $w=-1.09\pm0.37$ ($\pm68$\% confidence).   

Our new constraint that $\Ho=68.9\pm7.1$~\kmspmpc, when added
to the WMAP PDF, is shown in Fig.~\ref{figure:w_vs_Ho}
with dashed red contours.   This information for $\Ho$, which is
independent of $w$, improves the $w$-constraint by nearly a factor of two: $w=-0.98\pm0.20$.
This is probably about as accurate a result as one can obtain from measurements
of the water maser emission from \UGC\ with current equipment and a reasonable
amount of observing time.  Significantly improved measurements of this galaxy will 
likely await the completion of the Square Kilometer Array's high-frequency component.

Progress in the near future will come from measurements of other
megamaser galaxies.  The goal of the Megamaser Cosmology Project is to determine 
$\Ho$ with $\pm3$\% accuracy via $\approx10$ megamaser galaxy measurements.  
For example, were such measurements
to yield $\Ho=68.9\pm2.1$~\kmspmpc, this would further tighten the constraint on
the equation of state of dark energy (blue dotted contours in Fig.~\ref{figure:w_vs_Ho})  
to $w=-0.97\pm0.10$, independent of other methods.

\vskip 0.5truecm
We are grateful to the NRAO VLBA and GBT staff for their many contributions to 
the Megamaser Cosmology Project.
We thank the referee for many valuable comments on the original version of this paper.

\vskip 0.5truecm
\noindent
{\it Facilities:} \facility{VLBA, GBT, Effelsberg}

\pagebreak

\begin{deluxetable}{rrrrrrrrcc}
\tablecolumns{10} 
\tablewidth{0pc} 
\tablecaption{\UGC\ \HHO\ Maser Data}
\tabletypesize{\footnotesize}
\tablehead {
  \colhead{\Vlsr} && 
  \colhead{$\Theta_x$} & \colhead{$\sigma_{\Theta_x}$} &&
  \colhead{$\Theta_y$} & \colhead{$\sigma_{\Theta_y}$} &&
  \colhead{A} & \colhead{$\sigma_A$}
\\
  \colhead{(\kms)}  && 
  \colhead{(mas)} & \colhead{(mas)} &&
  \colhead{(mas)} & \colhead{(mas)} &&
  \colhead{(\kmsperyr)} &  \colhead{(\kmsperyr)}
            }
\startdata
    2472.8  && --0.183 &  0.009 && --0.190 &  0.009 && ...  & ...  \cr
    2477.9  && --0.179 &  0.008 && --0.197 &  0.021 && ...  & ...  \cr
    2562.9  && --0.116 &  0.005 && --0.176 &  0.028 && ...  & ...  \cr
    2564.6  && --0.119 &  0.004 && --0.153 &  0.005 && ...  & ...  \cr
    2566.3  && --0.122 &  0.003 && --0.149 &  0.003 && ...  & ...  \cr
    2568.0  && --0.123 &  0.003 && --0.143 &  0.009 &&  0.5 &  0.2 \cr
    2569.7  && --0.130 &  0.009 && --0.137 &  0.007 && ...  & ...  \cr
    2571.4  && --0.093 &  0.020 && --0.161 &  0.008 && ...  & ...  \cr
    2586.7  && --0.104 &  0.012 && --0.137 &  0.012 && ...  & ...  \cr
    2588.4  && --0.089 &  0.004 && --0.118 &  0.003 &&  0.0 &  0.2 \cr
    2590.1  && --0.084 &  0.009 && --0.102 &  0.005 && ...  & ...  \cr
    2591.8  && --0.087 &  0.008 && --0.103 &  0.006 && ...  & ...  \cr
    2593.5  && --0.096 &  0.004 && --0.107 &  0.004 &&  0.0 &  0.2 \cr
    2595.2  && --0.098 &  0.004 && --0.097 &  0.007 && ...  & ...  \cr
    2596.9  && --0.105 &  0.007 && --0.107 &  0.015 && ...  & ...  \cr
    2602.0  && --0.066 &  0.010 && --0.106 &  0.012 && ...  & ...  \cr
    2603.7  && --0.084 &  0.003 && --0.102 &  0.009 && ...  & ...  \cr
    2605.4  && --0.076 &  0.007 && --0.102 &  0.004 && ...  & ...  \cr
    2607.1  && --0.065 &  0.005 && --0.102 &  0.004 && ...  & ...  \cr
    2608.8  && --0.067 &  0.015 && --0.096 &  0.004 && ...  & ...  \cr
    2610.5  && --0.073 &  0.005 && --0.089 &  0.007 &&  0.0 &  0.2 \cr
    2612.2  && --0.077 &  0.003 && --0.087 &  0.007 &&  0.0 &  0.2 \cr
    2613.9  && --0.081 &  0.006 && --0.087 &  0.007 && ...  & ...  \cr
    2615.6  && --0.068 &  0.007 && --0.084 &  0.015 &&  0.1 &  0.2 \cr
    2617.3  && --0.072 &  0.007 && --0.085 &  0.007 && ...  & ...  \cr
    2629.2  && --0.045 &  0.012 && --0.079 &  0.013 && ...  & ...  \cr
    2630.9  && --0.042 &  0.009 && --0.054 &  0.010 && ...  & ...  \cr
    2632.6  && --0.049 &  0.004 && --0.059 &  0.004 && ...  & ...  \cr
    2634.3  && --0.050 &  0.004 && --0.077 &  0.004 && ...  & ...  \cr
    2636.0  && --0.064 &  0.008 && --0.063 &  0.005 && ...  & ...  \cr
    2668.3  && --0.009 &  0.006 && --0.034 &  0.007 && ...  & ...  \cr
    2670.0  && --0.020 &  0.006 && --0.032 &  0.005 && ...  & ...  \cr
    2678.5  && --0.006 &  0.004 && --0.024 &  0.004 && ...  & ...  \cr
    2680.2  && --0.016 &  0.003 && --0.011 &  0.003 && --0.1 &  0.2 \cr
    2681.9  && --0.008 &  0.002 && --0.008 &  0.002 && ...  & ...  \cr
    2683.6  && --0.008 &  0.002 && --0.008 &  0.002 &&  0.0 &  0.2 \cr
    2685.3  && --0.005 &  0.002 && --0.007 &  0.002 &&  0.1 &  0.2 \cr
\tablebreak
    2687.0  &&  0.000 &  0.002 &&  0.000 &  0.002 && ...  & ...  \cr
    2688.7  &&  0.000 &  0.002 &&  0.001 &  0.002 && --0.1 &  0.2 \cr
    2690.4  &&  0.005 &  0.002 &&  0.001 &  0.002 && ...  & ...  \cr
    2692.1  &&  0.000 &  0.003 &&  0.005 &  0.002 && --0.1 &  0.2 \cr
    2693.8  &&  0.009 &  0.002 &&  0.006 &  0.002 &&  0.0 &  0.2 \cr
    2695.5  &&  0.009 &  0.002 &&  0.010 &  0.003 && ...  & ...  \cr
    2697.2  &&  0.010 &  0.003 &&  0.007 &  0.004 && --0.1 &  0.2 \cr
    2698.9  &&  0.014 &  0.002 &&  0.010 &  0.002 &&  0.1 &  0.2 \cr
    2700.6  &&  0.012 &  0.002 &&  0.011 &  0.003 &&  0.1 &  0.2 \cr
    2702.3  &&  0.011 &  0.002 &&  0.011 &  0.003 && --0.1 &  0.2 \cr
    2704.0  &&  0.013 &  0.003 &&  0.011 &  0.004 && --0.2 &  0.2 \cr
    2705.7  &&  0.019 &  0.003 &&  0.026 &  0.003 && ...  & ...  \cr
    2707.4  &&  0.022 &  0.004 &&  0.028 &  0.010 && --0.3 &  0.2 \cr
    2709.1  &&  0.026 &  0.004 &&  0.016 &  0.014 && ...  & ...  \cr
    2710.8  &&  0.036 &  0.007 &&  0.033 &  0.007 && --0.4 &  0.2 \cr
    2712.5  &&  0.052 &  0.004 &&  0.029 &  0.007 && ...  & ...  \cr
    2714.2  &&  0.051 &  0.004 &&  0.048 &  0.005 && ...  & ...  \cr
    2715.9  &&  0.036 &  0.014 &&  0.046 &  0.005 && ...  & ...  \cr
    2717.6  &&  0.044 &  0.011 &&  0.048 &  0.005 && ...  & ...  \cr
    2719.3  &&  0.062 &  0.006 &&  0.054 &  0.009 && ...  & ...  \cr
    2721.0  &&  0.077 &  0.005 &&  0.060 &  0.006 &&  0.0 &  0.2 \cr
    2734.6  &&  0.046 &  0.003 &&  0.045 &  0.005 && ...  & ...  \cr
    2736.3  &&  0.040 &  0.006 &&  0.042 &  0.012 &&  0.2 &  0.2 \cr
    2738.0  &&  0.045 &  0.008 &&  0.054 &  0.006 && ...  & ...  \cr
    2739.7  &&  0.044 &  0.008 &&  0.042 &  0.003 && --0.1 &  0.2 \cr
    2741.4  &&  0.050 &  0.004 &&  0.051 &  0.004 && --0.1 &  0.2 \cr
    2743.1  &&  0.049 &  0.005 &&  0.053 &  0.004 && --0.1 &  0.2 \cr
    2744.8  &&  0.053 &  0.003 &&  0.048 &  0.006 && --0.2 &  0.2 \cr
    2746.5  &&  0.048 &  0.009 &&  0.047 &  0.003 && ...  & ...  \cr
    2748.2  &&  0.058 &  0.004 &&  0.050 &  0.005 && --0.4 &  0.2 \cr
    2749.9  &&  0.058 &  0.004 &&  0.052 &  0.014 && ...  & ...  \cr
\tablebreak
    3234.4  && --0.403 &  0.010 && --0.406 &  0.010 &&  1.6 &  0.2 \cr
    3239.5  && --0.401 &  0.012 && --0.429 &  0.005 &&  1.3 &  0.2 \cr
    3241.2  && --0.404 &  0.005 && --0.418 &  0.005 &&  1.8 &  0.2 \cr
    3242.9  && --0.393 &  0.004 && --0.433 &  0.004 &&  2.5 &  0.2 \cr
    3244.6  && --0.389 &  0.003 && --0.430 &  0.003 &&  1.1 &  0.2 \cr
    3246.3  && --0.397 &  0.003 && --0.433 &  0.004 &&  1.9 &  0.2 \cr
    3248.0  && --0.397 &  0.004 && --0.431 &  0.009 &&  1.8 &  0.2 \cr
    3249.7  && --0.405 &  0.004 && --0.438 &  0.008 &&  1.6 &  0.2 \cr
    3251.4  && --0.405 &  0.003 && --0.439 &  0.005 &&  1.2 &  0.2 \cr
    3253.1  && --0.401 &  0.004 && --0.437 &  0.005 &&  1.0 &  0.2 \cr
    3254.8  && --0.413 &  0.007 && --0.414 &  0.009 &&  1.6 &  0.2 \cr
    3256.5  && --0.422 &  0.006 && --0.442 &  0.011 &&  0.9 &  0.2 \cr
    3258.2  && --0.411 &  0.004 && --0.443 &  0.004 &&  1.1 &  0.2 \cr
    3259.9  && --0.407 &  0.003 && --0.442 &  0.008 &&  1.5 &  0.2 \cr
    3261.6  && --0.406 &  0.004 && --0.445 &  0.005 &&  1.8 &  0.2 \cr
    3263.3  && --0.403 &  0.003 && --0.463 &  0.015 &&  1.9 &  0.2 \cr
    3266.7  && --0.401 &  0.009 && --0.454 &  0.009 &&  1.7 &  0.2 \cr
    3268.4  && --0.412 &  0.011 && --0.484 &  0.007 &&  1.9 &  0.2 \cr
    3270.1  && --0.402 &  0.006 && --0.470 &  0.015 &&  2.1 &  0.2 \cr
    3271.8  && --0.403 &  0.006 && --0.478 &  0.006 &&  1.9 &  0.2 \cr
    3273.5  && --0.404 &  0.006 && --0.477 &  0.006 &&  4.2 &  0.2 \cr
    3287.1  && --0.410 &  0.017 && --0.482 &  0.008 &&  2.5 &  0.8 \cr
    3288.8  && --0.406 &  0.004 && --0.474 &  0.004 &&  6.6 &  0.8 \cr
    3290.5  && --0.404 &  0.005 && --0.506 &  0.005 &&  6.1 &  0.8 \cr
    3292.2  && --0.409 &  0.003 && --0.485 &  0.009 &&  7.3 &  0.8 \cr
    3293.9  && --0.404 &  0.003 && --0.487 &  0.003 &&  8.4 &  0.8 \cr
    3295.6  && --0.403 &  0.003 && --0.490 &  0.004 &&  7.4 &  0.8 \cr
    3297.3  && --0.403 &  0.002 && --0.492 &  0.006 &&  7.2 &  0.8 \cr
    3299.0  && --0.405 &  0.002 && --0.489 &  0.003 &&  5.9 &  0.8 \cr
    3300.7  && --0.406 &  0.002 && --0.489 &  0.004 &&  5.7 &  0.8 \cr
    3302.4  && --0.409 &  0.002 && --0.490 &  0.004 &&  4.8 &  0.8 \cr
    3304.1  && --0.406 &  0.002 && --0.493 &  0.002 &&  4.7 &  0.8 \cr
    3305.8  && --0.404 &  0.002 && --0.495 &  0.005 &&  5.0 &  0.8 \cr
    3307.5  && --0.406 &  0.003 && --0.494 &  0.005 &&  5.4 &  0.8 \cr
    3309.2  && --0.404 &  0.002 && --0.491 &  0.004 &&  4.9 &  0.8 \cr
    3310.9  && --0.411 &  0.003 && --0.493 &  0.003 &&  5.9 &  0.8 \cr
    3312.6  && --0.410 &  0.003 && --0.499 &  0.004 &&  6.0 &  0.8 \cr
    3314.3  && --0.399 &  0.005 && --0.503 &  0.005 &&  5.3 &  0.8 \cr
    3316.0  && --0.412 &  0.010 && --0.502 &  0.004 &&  5.3 &  0.8 \cr
    3317.7  && --0.405 &  0.019 && --0.499 &  0.005 &&  5.5 &  0.8 \cr
    3319.4  && --0.418 &  0.015 && --0.518 &  0.028 &&  5.9 &  0.8 \cr
    3358.5  && --0.437 &  0.007 && --0.511 &  0.006 &&  2.4 &  0.2 \cr
\tablebreak
    3660.7  && --1.198 &  0.010 && --1.517 &  0.012 &&  0.2 &  0.2 \cr
    3735.5  && --0.977 &  0.005 && --1.139 &  0.008 && ...  & ...  \cr
    3737.2  && --0.980 &  0.006 && --1.133 &  0.004 &&  0.1 &  0.2 \cr
    3738.9  && --0.992 &  0.012 && --1.119 &  0.011 && ...  & ...  \cr
    3740.6  && --0.971 &  0.005 && --1.157 &  0.005 &&  0.2 &  0.2 \cr
    3745.7  && --0.965 &  0.014 && --1.143 &  0.020 &&  0.2 &  0.2 \cr
    3747.4  && --0.973 &  0.006 && --1.123 &  0.009 && ...  & ...  \cr
    3749.1  && --0.981 &  0.009 && --1.109 &  0.005 && --0.1 &  0.2 \cr
    3750.8  && --0.988 &  0.007 && --1.095 &  0.007 && ...  & ...  \cr
    3752.5  && --0.965 &  0.011 && --1.111 &  0.005 && ...  & ...  \cr
    3754.2  && --0.968 &  0.015 && --1.094 &  0.008 && --0.2 &  0.2 \cr
    3755.9  && --0.940 &  0.008 && --1.092 &  0.004 && ...  & ...  \cr
    3757.6  && --0.953 &  0.006 && --1.092 &  0.005 &&  0.0 &  0.2 \cr
    3759.3  && --0.945 &  0.003 && --1.104 &  0.004 && ...  & ...  \cr
    3761.0  && --0.944 &  0.004 && --1.101 &  0.004 &&  0.1 &  0.2 \cr
    3764.4  && --0.944 &  0.014 && --1.082 &  0.015 && ...  & ...  \cr
    3808.6  && --0.782 &  0.006 && --0.833 &  0.006 && ...  & ...  \cr
    3837.5  && --0.830 &  0.010 && --0.917 &  0.028 && ...  & ...  \cr
    3839.2  && --0.811 &  0.009 && --0.905 &  0.011 &&  0.2 &  0.2 \cr
    3840.9  && --0.814 &  0.010 && --0.914 &  0.005 && ...  & ...  \cr
    3842.6  && --0.832 &  0.006 && --0.916 &  0.007 && ...  & ...  \cr
    3846.0  && --0.766 &  0.012 && --0.874 &  0.014 && ...  & ...  \cr
    3851.1  && --0.791 &  0.007 && --0.858 &  0.029 && ...  & ...  \cr
    3859.6  && --0.760 &  0.004 && --0.848 &  0.005 && ...  & ...  \cr
    3861.3  && --0.764 &  0.004 && --0.866 &  0.004 && ...  & ...  \cr
    3863.0  && --0.764 &  0.004 && --0.882 &  0.004 && ...  & ...  \cr
    3864.7  && --0.760 &  0.007 && --0.874 &  0.005 && ...  & ...  \cr
    3866.4  && --0.761 &  0.009 && --0.868 &  0.007 &&  0.3 &  0.2 \cr
    3868.1  && --0.765 &  0.010 && --0.870 &  0.004 && ...  & ...  \cr
    3871.5  && --0.751 &  0.030 && --0.874 &  0.021 && ...  & ...  \cr
    3878.3  && --0.761 &  0.016 && --0.845 &  0.023 && ...  & ...  \cr
    3880.0  && --0.758 &  0.004 && --0.871 &  0.004 && ...  & ...  \cr
    3881.7  && --0.750 &  0.004 && --0.863 &  0.003 &&  0.5 &  0.2 \cr
    3883.4  && --0.753 &  0.003 && --0.854 &  0.009 && ...  & ...  \cr
    3885.1  && --0.746 &  0.003 && --0.854 &  0.013 &&  0.0 &  0.2 \cr
    3886.8  && --0.750 &  0.007 && --0.865 &  0.005 && ...  & ...  \cr
    3890.2  && --0.758 &  0.024 && --0.849 &  0.017 && ...  & ...  \cr
    3891.9  && --0.744 &  0.008 && --0.857 &  0.007 && ...  & ...  \cr
    3893.6  && --0.715 &  0.006 && --0.856 &  0.007 && ...  & ...  \cr
    3903.8  && --0.738 &  0.007 && --0.830 &  0.005 && ...  & ...  \cr
    3905.5  && --0.727 &  0.005 && --0.838 &  0.011 && ...  & ...  \cr
    3907.2  && --0.733 &  0.004 && --0.835 &  0.008 && ...  & ...  \cr
    3908.9  && --0.727 &  0.004 && --0.833 &  0.009 && ...  & ...  \cr
    3910.6  && --0.720 &  0.003 && --0.827 &  0.004 && ...  & ...  \cr
    3912.3  && --0.730 &  0.007 && --0.824 &  0.007 && ...  & ...  \cr
    3914.0  && --0.733 &  0.006 && --0.809 &  0.014 && ...  & ...  \cr
\enddata
\tablecomments {} 
\label{table:data}
\end{deluxetable}

\begin{deluxetable}{lrrrrrrrrrr}
\tablecolumns{11} 
\tablewidth{0pc} 
\tablecaption{Parameter Correlation Coefficients}
\tablehead {
  \colhead{}  
  &\colhead{\Ho} &\colhead{$M$} &\colhead{$V_0$} &\colhead{$x_0$} &\colhead{$y_0$} 
  &\colhead{$i$} &\colhead{$\didr$} &\colhead{$p$} &\colhead{$\dpdr$} &\colhead{$V_p$} 
            }
\startdata
  $\Ho$     &1.000 &$-$0.846  &0.024  &0.029  &0.061  &0.318  &0.084  &0.034 &$-$0.027  &0.410\cr
  $M$      &$-$0.846 &1.000   &0.024 &$-$0.021 &$-$0.071 &$-$0.432 &$-$0.157 &$-$0.019  &0.027  &0.128\cr
  $V_0$     &0.024 &0.024  &1.000 &$-$0.011  &0.053 &$-$0.018 &$-$0.254  &0.314 &$-$0.003  &0.073\cr
  $x_0$     &0.029 &$-$0.021 &$-$0.011  &1.000  &0.162 &$-$0.371  &0.029 &$-$0.217  &0.215  &0.023\cr
  $y_0$     &0.061 &$-$0.071  &0.053  &0.162  &1.000  &0.193 &$-$0.337  &0.183 &$-$0.268 &$-$0.013\cr
  $i$       &0.318 &$-$0.432 &$-$0.018 &$-$0.371  &0.193  &1.000  &0.016  &0.213 &$-$0.196 &$-$0.153\cr
 $\didr$    &0.084  &$-$0.157 &$-$0.254  &0.029 &$-$0.337  &0.016  &1.000 &$-$0.086  &0.179 &$-$0.112\cr
  $p$       &0.034 &$-$0.019  &0.314 &$-$0.217  &0.183  &0.213 &$-$0.086  &1.000 &$-$0.433  &0.023\cr
 $\dpdr$    &$-$0.027  &0.027 &$-$0.003  &0.215 &$-$0.268 &$-$0.196  &0.179 &$-$0.433  &1.000 &$-$0.007\cr
  $V_p$     &0.410  &0.128 &0.073  &0.023 &$-$0.013 &$-$0.153 &$-$0.112  &0.023 &$-$0.007  &1.000\cr
\enddata
\tablecomments {\footnotesize
   Pearson product-moment correlation coefficients calculated from $10^7$ McMC trial parameter
   values.  Parameter definitions are given in the text and the notes in 
   Table~\ref{table:fitted_parameters} 
              }
\label{table:correlations}
\end{deluxetable}

\begin{deluxetable}{ccccl}
\tablecolumns{5} 
\tablewidth{0pc} 
\tablecaption{\UGC\ Basic Disk Model}
\tablehead {
  \colhead{Parameter} && 
  \colhead{Priors} & \colhead{Posterioris} &\colhead{Units} 
            }
\startdata
  $\Ho$      && $...$         & $68.9\pm7.1$    & \kmspmpc\      \cr
  $V_0$      && $...$         & $3320\pm1$      & \kms\          \cr
  $V_p$      && $151\pm163$   & $146\pm175$     & \kms\          \cr
  $M$        && $...$         & $1.16\pm0.12$   & $10^7$ \Msun\  \cr
  $x_0$      && $...$         & $-0.402\pm0.002$& mas            \cr
  $y_0$      && $...$         & $-0.460\pm0.002$& mas            \cr
 $i(r_{ref})$&& $...$         & $90.6\pm0.4$    & deg            \cr
 $\didr$     && $...$         & $-7.6\pm2.0$    & deg mas$^{-1}$ \cr
 $p(r_{ref})$&& $...$         & $221.5\pm0.2$   & deg            \cr
 $\dpdr$     && $...$         & $-2.0\pm1.1$    & deg mas$^{-1}$ \cr
  $e$        && $0$           & $0$             & ...            \cr
$\omega$     && $0$           & $0$             & deg            \cr
$\domegadr$  && $0$           & $0$             & deg mas$^{-1}$ \cr
\enddata
\tablecomments {\footnotesize Parameters are as follows: Hubble constant (\Ho),
   observed velocity of the central black hole $V_0$ (non-relativistic, 
   optical definition in CMB frame), peculiar velocity $V_p$ with respect to 
   Hubble flow in cosmic microwave background frame (\ie\ $cz = V_0 + V_p$), 
   black hole mass ($M$), eastward ($x_0$) and northward ($y_0$) 
   position of black hole with respect to the reference masers 
   ($2684< V_{LSR} <2693$ \kms),
   disk inclination $i(r_{ref})$ at the reference angular radius of 0.60 mas, 
   and inclination warping (change of inclination with radius: \didr),
   disk position angle $p(r_{ref})$ at the reference angular radius 
   and position angle warping (change of position position angle with radius: \dpdr),
   gas orbital eccentricity ($e$) and angle of pericenter with respect
   to the line of sight ($\omega$) and its derivative with angular radius (\domegadr).  
   Flat priors were used, except where listed.  Posteriori values are
   medians of their marginalized probability density functions and uncertainties
   give $\pm34$\% confidence ranges, scaled by the square-root of the (reduced)
   chi-squared per degree of freedom (except for $V_p$ which is controlled
   by its prior).
              }
\label{table:fitted_parameters}
\end{deluxetable}

\end{document}